\documentclass{elsart}
\usepackage{epsfig}

\usepackage{amssymb}

\begin{document}

\begin{frontmatter}

% use the thanksref command within \title, \author or \address for footnotes:
% \author{Name\thanksref{label2}}
% including your email address:
% \address{Address\thanksref{email}}
% \thanks[email]{E-mail: }

\title{Sex-oriented stable matchings of the Marriage Problem with correlated and incomplete information}

% use optional labels to link authos explicitly to addresses:
\author[roma]{Guido Caldarelli\thanksref{eGC}}, 
\author[unifr]{Andrea Capocci\thanksref{eAC}} and 
\author[unifr]{Paolo Laureti\thanksref{ePL}}
\address[roma]{INFM Unit\`{a} ROMA1 and Dip. di Fisica - Universit\`{a} di Roma "La Sapienza", P.le A.Moro 2, 00185 Roma, Italy} 
\address[unifr]{Institut de Physique Th{\'e}orique, Universit{\'e} de Fribourg, Perolles CH-1700, Switzerland}

\thanks[eGC]{gcalda@pil.phys.uniroma1.it}
\thanks[eAC]{andrea.capocci@unifr.ch}
\thanks[ePL]{paolo.laureti@unifr.ch}
\date{\today}

\begin{abstract}
In the Stable Marriage Problem two sets of agents must be paired
according to mutual preferences, which may happen to conflict. We
present two generalizations of its sex-oriented version, aiming to
take into account correlations between the preferences of agents
and costly information. Their effects are investigated both
numerically and analytically.
\end{abstract}

\end{frontmatter}

Game theory \cite{Fud} has become a fruitful source of inspiration
for both physicists and economists in recent times. The idea of
Nash equilibria \cite{Nash}, as opposed to the minimisation of an
Hamiltonian, is a powerful tool to investigate the behaviour of
selfish agents aiming to maximise their individual utility. A Nash
Equilibrium is a state in which any variation of an agent's
strategy results in a worse performance for herself, the other
agents' strategies being unchanged. A growing effort is being
devoted to analyzing situations in which only a subset of possible
strategies can be explored, due to the limited capabilities of
real agents \cite{Sim}, that is a source of market inefficiency
\cite{Zha}. The Stable Marriage Problem provides a simple and
natural environment to represent agents with bounded rationality.

The Stable Marriage Problem \cite{Gus} describes a system where
two classes of $N$ players (e.g., {\it men} and {\it women}) have
to be matched pairwise. Each player is been assigned his/her list
of preferred partners. In the original model, the lists are drawn
at random and are independent from one another. If man $m$ marries
woman $w$ we attribute him an energy equal to the ranking of $w$
in $m$'s list. This way each player tends to minimise his/her
energy, or unhappiness. Let us define the matrices $f$ (for women) and
$h$ (for men), such that the element $f(w,m)$ denotes the rank of
man $m$ in $w$'s list and $h(m,w)$ the rank of woman $w$ in $m$'s
list. We call {\em stable state} a collection of couples
$\mathcal{M} =\{(m,w)_i\}_{i=1,\ldots,N}$ where there is no man
$m_i$ and no woman $w_j$, members of two different couples ($i$
and $j$), who would both prefer to marry each other rather than
staying with their respective partners. The energies per person
for women ($f$) and men ($h$) in a given matching are defined by
\begin{equation}
\epsilon_{f}=\frac{1}{N}\sum_{w=1}^{N}f(w,m)
\end{equation}
and
\begin{equation}
\epsilon_{f}=\frac{1}{N}\sum_{m=1}^{N}h(m,w),
\end{equation}
and we will consider their averages over many instances of the
preference lists.

Many numerical studies are based on the Gale-Shapley Algorithm
(GSA) \cite{GS}. According to the GSA, the agents of one class
propose to marry to the agents of the other class, who judge
whether to accept or not. This algorithm finds the stable state
with the minimal proponents' energy.
In the man-oriented GSA, man $m$ makes a proposal to the first
woman on his list. If she accepts, they get engaged; if she
refuses $m$ goes on proposing to the next woman on his list.
Conversely, woman $w$ accepts a proposal if she is not engaged or
if she is engaged with a man $m'$ worse ranked (in her list) than
the proponent. In this case, $m'$ restarts proposing to the
following women in his list.

When all men have run through their lists and all women are
engaged, the algorithm stops and the resulting matching
$\mathcal{M}_{h}$ is stable. Of course, the algorithm could also
be run reversing the roles and proponents are always better
rewarded. Indeed, it is possible to derive a general relation
between the energies of women and men in a stable state. Through a
mean-field approach, it has been shown in \cite{Om} and \cite{Dz}
that the energy of men is
\begin{equation}
\epsilon_{h} = \log N + 0.5772
\end{equation}
and that the energy of women $\epsilon_{f}$ can be determined by
the relation $\epsilon_f = N/\epsilon_h$, which holds for all
stable matchings.

A more general case is represented in \cite{cc} where the
preference lists are correlated in such a way that certain agents
rank, on average, higher than others. This amounts to provide them
with an intrinsic characteristic, which we shall call {\em
beauty}. Lists are built as follows: for each man $m$, woman $w$
is assigned a score $\mu_{m}+ U·\mu_{w}$, where $\mu_{m}$ and
$\mu_{w}$ are random variables uniformly distributed in the range
$[0,1]$ and $U$ is a parameter tuning the effect of beauty. The
wish list of player $m$ results from ordering all the scores thus
obtained. Setting $U=0$, one recovers the original uncorrelated
model. One observes \cite{cc} that, as $U$ is increased, the
average total energy grows. Moreover, the energy gap between
proponents and judges decreases, while an inequality is introduced
in the energies of "beautiful" and "ugly" agents, regardless their
role.

The Stable Marriage Problem has been widely studied as a model of
economical systems, since it mimics agents maximising their
individual utility in a competing environment. As pointed out by
behavioural economists \cite{Sim}, the cost of information has to
be taken into account. Since agents can exert a finite effort, the
exploration of all possible alternatives is bounded.
The effects of limited information in matching problems are investigated
in a following paper \cite{Zha2} with a more general approach.

In our sex-oriented marriage problem, one can bound the rationality of agents
by arranging them on a regular lattice of linear dimension $L$ and
assuming that they only know partners within a given euclidean
distance $d$. Unreachable individuals (who may happen to be the
best ranked ones) figure as holes in the wish lists. Under this
assumption, each agent only knows a fraction $\alpha \sim
(\frac{d}{L})^{K}$ of the total number of agents, where $K$ is the
lattice dimension.

\begin{figure}
%\narrowtext
%\centerline{\epsfxsize=3.375in
\centerline{\epsfxsize=5.in \epsffile{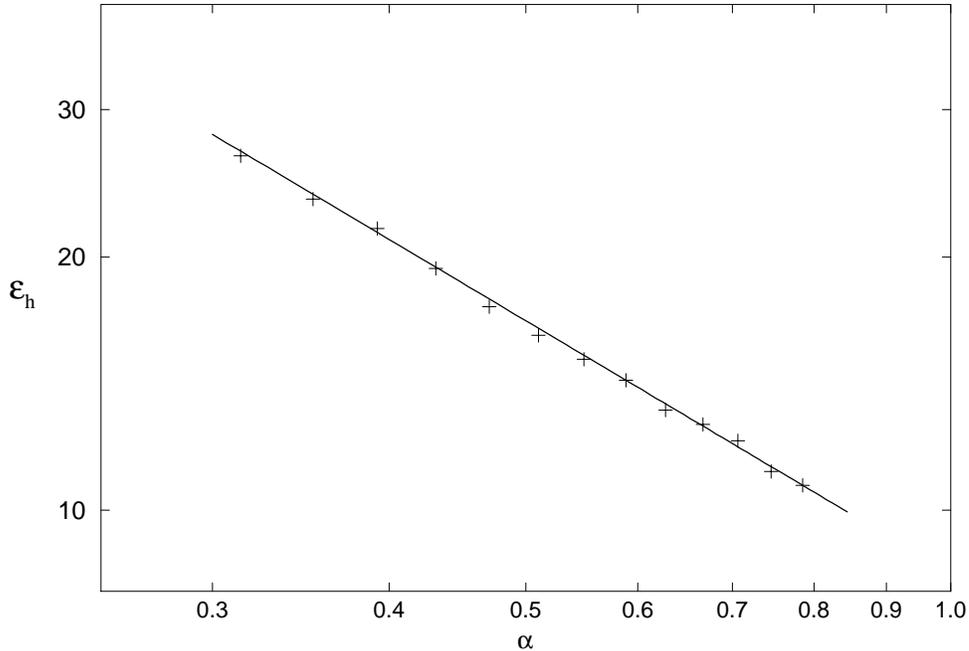} }

\caption{Energy of men averaged over $200$ realizations as 
a function of $\alpha$ in a bidimensional model with $N=2500$ plotted on log-log scale. 
``Plus'' symbols correspond to simulation data. The solid line is obtained from eq. (\ref{eHa2}).} \label{fig1}
\end{figure}

\begin{figure}
%\narrowtext
%\centerline{\epsfxsize=3.375in
\centerline{\epsfxsize=5.in \epsffile{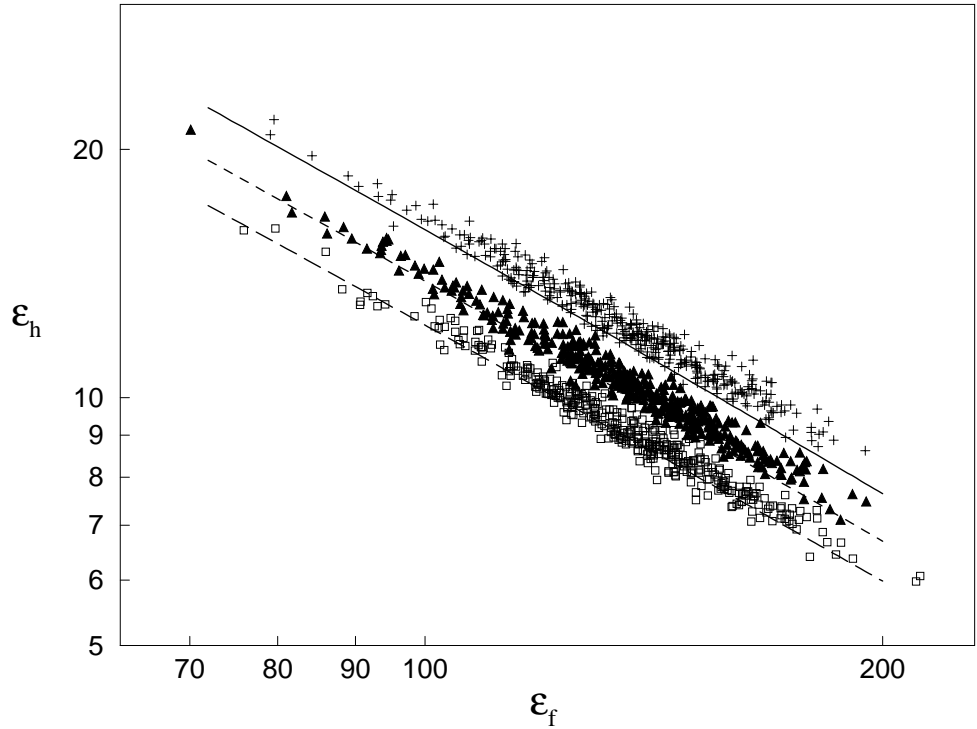} }

\caption{Energy of men and women in stable matchings in a one
dimensional model for different values of $\alpha$ plotted on log-log scale.
Symbols correspond to simulation data, lines are obtained from eq.
(\ref{eHa1}).
The solid line and ``plus'' symbols refer to $\alpha = 0.6$; the dashed line and triangles refer to $\alpha = 0.7$; the long-dashed line and the squares refer to $\alpha = 0.8$} \label{fig2}
\end{figure}

In a GSA with "full" lists, the energy of a man equals the number
of proposals needed to get married. If lists are "sparse", each
subsequent proposal results in adding $\frac{1}{\alpha}$, on
average, to his energy. The energy of men, in this case, equals
the number of proposals divided by $\alpha$. If we follow the same
procedure of \cite{Om}, neglecting correlations between the
position of the holes, we find
\begin{equation} \label{eHa1}
\epsilon_{h}(\alpha) = \frac{\alpha
\epsilon_{f}(\alpha)}{N[1-(1-\frac{\epsilon_{f}(\alpha)}{N})^{\alpha}]^2}
\end{equation}
where
\begin{equation} \label{eHa2}
\epsilon_H(\alpha) = \frac{1}{\alpha}(\log N + 0.5772)
\end{equation}
These relations are verified by numerical simulations, as shown in
figure \ref{fig1} and \ref{fig2}.

We investigated both numerically and analytically a generalised
Stable Marriage Problem. First we assigned an individual {\it
beauty} to the agents, which results in correlated preference
lists. This correlation smoothes the favorable edge of proponents
but introduces an energy gap related to the intrinsic feature of
agents. Then, we studied a model where agents have an incomplete
information about potential partners. We show that a richer
information implies a greater competition, but also a lower
average energy.

% main text
%\section{}
%\label{}

\end{document}